\documentstyle[11pt,cs11,html,epsf]{article}

\begin{document}

\title{The late-type stellar component in the ROSAT All-Sky Survey at
 high galactic latitude}

\author{F.-J. Zickgraf}\affil{Hamburger Sternwarte, Gojenbergsweg 112, 
D-21029 Hamburg, Germany}
\author{J.M. Alcal\'a and E. Covino}
\affil{Osservatorio Astronomico di Capodimonte, I-80131 Napoli, Italy}
\author{J. Krautter and I. Appenzeller}
\affil{Landessternwarte K\"onigstuhl, D-69117 Heidelberg, Germany}
\author{S. Frink}
\affil{University of California San Diego, La Jolla, CA 92093-0424, USA}
\author{M.F. Sterzik}
\affil{European Southern Observatory, Santiago 19, Chile}







\begin{abstract}
We investigated the 
properties of the late-type stellar component
in the RASS at high-galactic latitude $|b|$ based on an optically 
identified sample of ROSAT All-Sky Survey (RASS)
X-ray sources. The stellar sample comprises 
$\sim250$ objects in six study areas covering 685 deg$^2$ at $|b|> 20\deg$.
We spectroscopically detected a
significant fraction of lithium-rich pre-main sequence (PMS) objects 
including even M-type stars. In an area located about 20$\deg$
south of the Tau-Aur star formation region (SFR) and close to the Gould Belt, 
we found a large 
fraction of 40\% PMS stars among the K-type stellar
counterparts. In other areas we found a smaller 
but still significant fraction of Li-rich stars.  
We compared the $\log N - \log S$ distribution with published galactic 
distribution models for different age groups and with results from the ROSAT
Galactic Plane Survey. For the sample south of Tau-Aur we find an excess of 
PMS stars compared to model calculations while in the other areas the observed 
$\log N - \log S$ is close to the model predictions.
We started to investigate the proper motions and radial velocities of both, 
the young lithium-rich and the older stellar counterparts. Radial velocities 
and proper motions of the Li-rich stars in the area south of Tau-Aur are 
consistent with membership to the Tau-Aur SFR. The non-PMS stars show a 
wider spread in  radial velocities and proper motions.

\end{abstract}




%
%
%


\section{High galactic latitude sample}

\begin{figure}
\plotfiddle{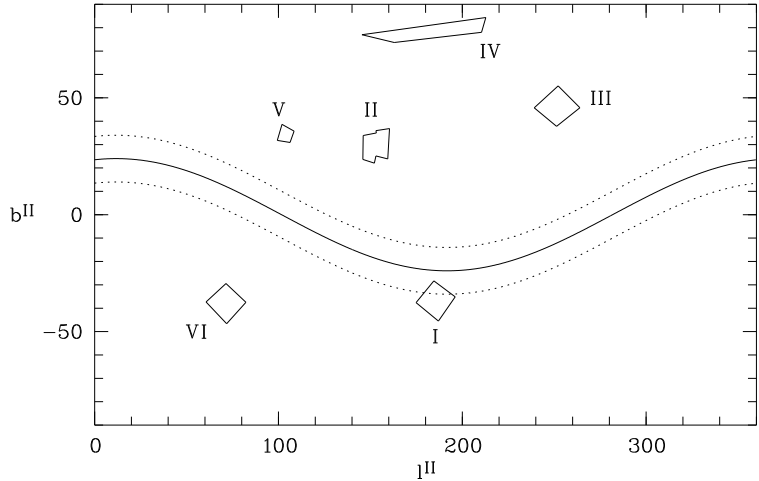}{6.5cm}{-90}{40}{40}{-170}{210}
\caption{Distribution of the study areas in galactic coordinates. The solid 
and dashed lines denote the location of the Gould Belt as given by Guillout 
et al. (1998).} 
\label{area}
\end{figure}

The optical identification of a complete sample of 674 RASS X-ray sources 
yielded a stellar subsample of 274 sources (Zickgraf et al. 1997, Appenzeller 
et al. 1998, Krautter et al. 1999). The RASS sample is distributed over six 
study areas located at $|b| > 20\deg$ and covering 685 sq. degrees (see 
Fig. \ref{area}). The 
flux limit is mostly 1.8\,10$^{-13}$\,erg\,s$^{-1}$\,cm$^{-2}$ and in the 
smaller area V 6\,10$^{-14}$\,erg\,s$^{-1}$\,cm$^{-2}$. The stellar sample 
contains about 20\% F and as much G stars, 33\% K stars (with and without 
H$\alpha$ emission),
and 21\% M stars, most of them, i.e. 90\%, of type dMe. 

\section{$\log N - \log S$ distribution}

\begin{figure}
\plotfiddle{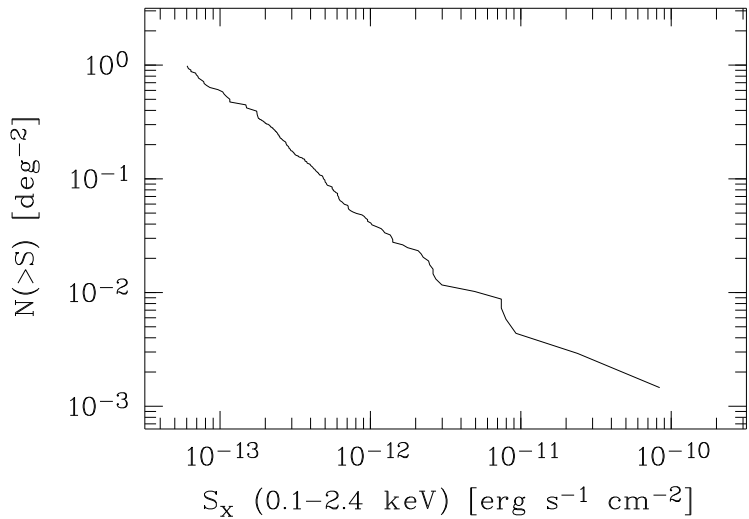}{6.5cm}{-90}{40}{40}{-170}{210}
\caption{Combined area corrected $\log N - \log S$ distribution of stellar counterparts (excluding cataclysmic variables and white dwarfs).} 
\label{logns}
\end{figure}

The $\log N - \log S$ distribution of the stellar sample (excluding 
cataclysmic variables and white dwarfs) is shown in Fig. \ref{logns}. In 
the flux range $S_{\rm x} = 2\,10^{-13} - 
1\,10^{-11}$\,erg\,s$^{-1}$\,cm$^{-2}$ the slope 
is $-1.36\pm0.11$ calculated using a 
maxi\-mum-like\-li\-hood me\-thod. Below 2\,10$^{-13}$\,erg\,s$^{-1}$\,cm$^{-2}$ the 
distribution is even flatter with a slope of $-1.2$. The 
slope of the distribution is therefore flatter than the Euclidean slope of 
$-1.5$. It is also flatter than  derived for low $|b|$ from the Galactic Plane
Survey by Motch et al. (1997) who obtained a slope of 
$-1.48\pm0.23$ for an area in Cygnus. The deviation from the  Euclidean value 
indicates that the 
stellar sample at high $|b|$ is affected by the scale height of the galactic 
distribution of stellar X-ray emitters. While the average 
$\log N - \log S$ is consistent with age dependent models for the space 
distribution of X-ray active stars by Guillout et al. (1996) for $|b| =
30^{\circ}$, area I (south of Tau-Aur) deviates as discussed already by 
Zickgraf et al. (1998) due to an excess of lithium-rich PMS stars 
(see Sect. \ref{lithium}). 

\section{Lithium-rich stars}
\label{lithium}
\index{lithium}
During several observing runs spectra of $\sim70$ G, K, and M stars  have been 
obtained (medium and high resolution) in order to search for the 
\ion{Li}{1 $\lambda$6708}\,\AA\
absorption line. The results are summarized in Tab.~\ref{litab}. 
Equivalent widths have been converted to abundances using the curves of 
growth by Pavlenko \& Magazz\'u (1996). So far, in 15 of 67 K stars 
strong lithium lines have been found. Most of them, i.e. eight 
K stars,  are located in area I which contains 20 K and Ke stars. 
In areas II, III, and VI 7 of 47 K and Ke stars exhibit strong Li-absorption 
lines. Even 2 of 40 M stars show very strong 
lithium lines, one in area I and the other  one in area II. 
Whereas the survey in 4 of the 6  areas is 
complete for the K and M stars, it is not yet complete for the G stars. 
However, three G stars in area I exhibit strong lithium, and three further 
G stars have considerable Li-absorption. In area VI  one of 
three G stars observed so far exhibits considerable Li-absorption.

\section{Proper motions and radial velocities}
\begin{figure}[tbh]
\plottwo{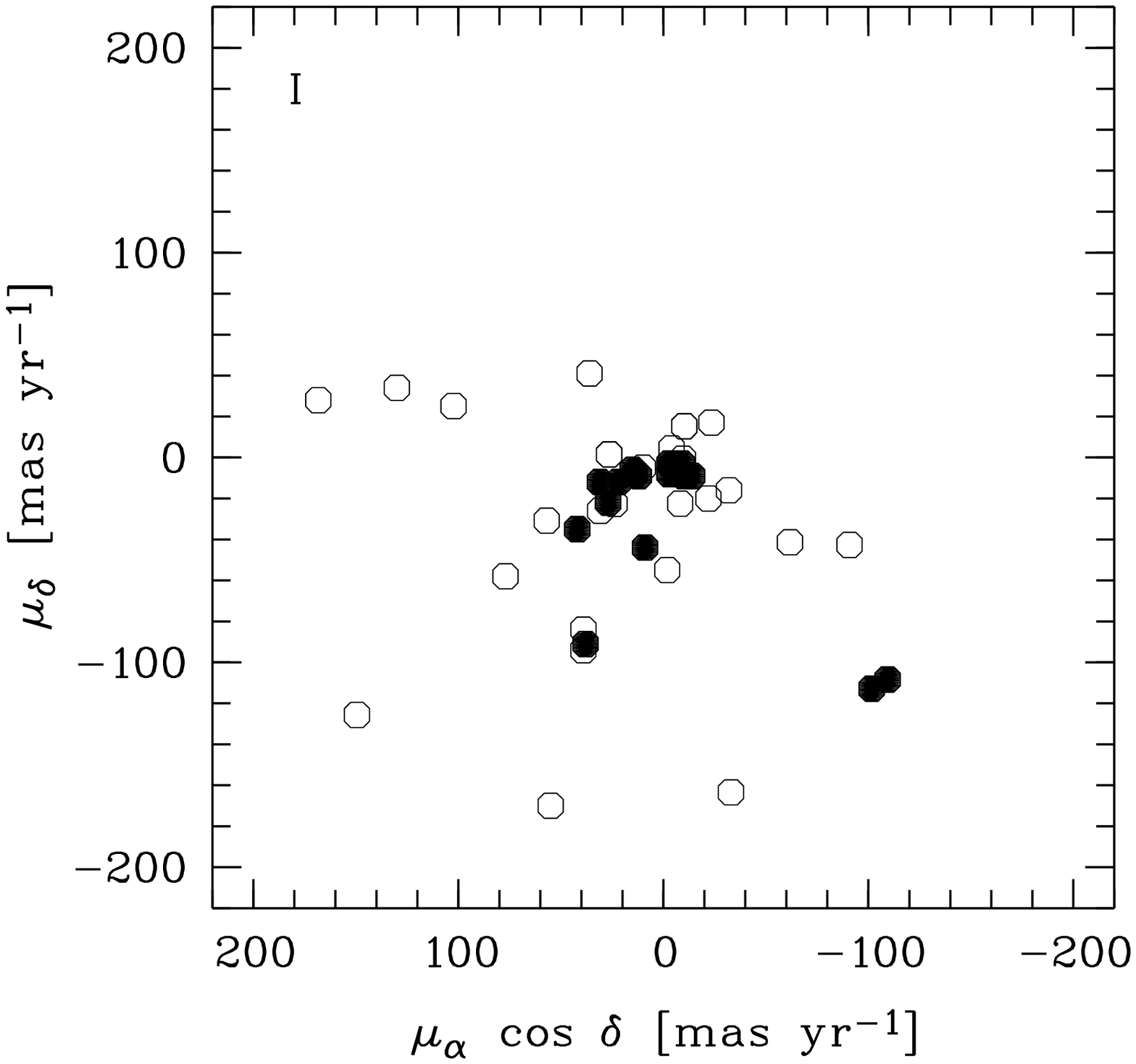}{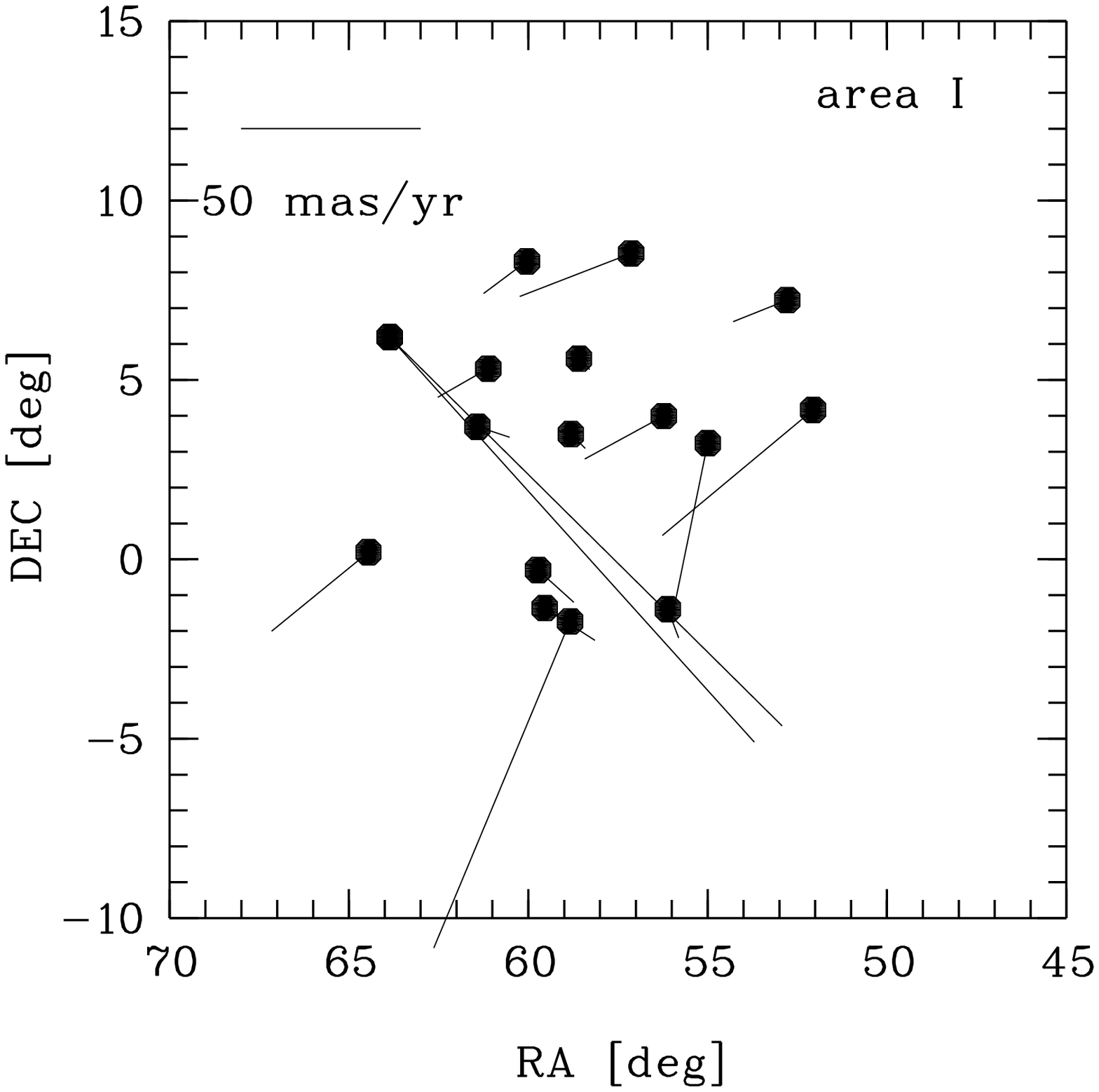}
\caption[]{Proper motions (left panel) and projected directions of 
motion (right panel) 
for stellar X-ray emitters in area I. Filled symbols: Li-rich stars. Open 
symbols: stars with $W$(Li) $< 50$\,m\AA .
}
\label{prop1}
\end{figure}

\begin{figure}[tbh]
\plotfiddle{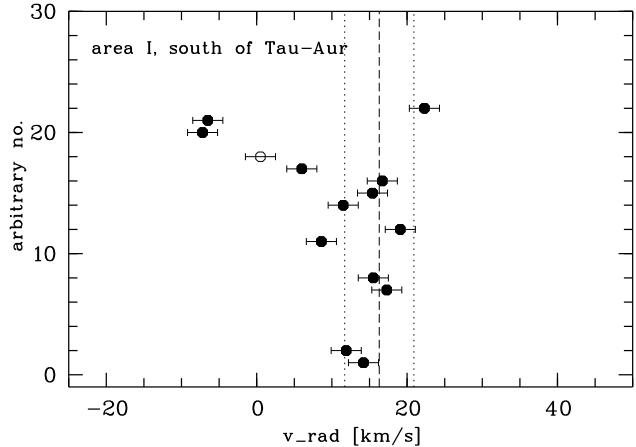}{5.0cm}{0}{50}{50}{-150}{-45}
\caption[]{Heliocentric radial velocities for area I. Filled symbols: lithium-rich stars, open symbols: stars without lithium. The velocity 
range of Tau-Aur is marked by long and short dashed lines. 
}
\label{radvel}
\end{figure}

\index{proper motion}
\index{radial velocity}
We searched the HIPPARCOS, TYCHO, PPM, STARNET (R\"oser 1996), and ACT 
(Urban et al. 1997) 
catalogs for the stars in our sample in order to obtain proper motions. 
Figure \ref{prop1} shows the proper motions and the respective vectors of 
motion for area I. The Li-rich stars have proper motions 
clustering in a smaller range than stars without significant lithium.
The two lithium-rich stars in the lower right corner of the left panel of
Fig.~\ref{prop1} deviating in proper motion are V774\,Tau and V891\,Tau. 
They also have a different direction 
of motion and different radial velocities (cf. Fig. \ref{radvel}) and possibly
belong to the UMa cluster at a distance of about 20\,pc. 

\begin{figure}[tbh]
{\plottwo{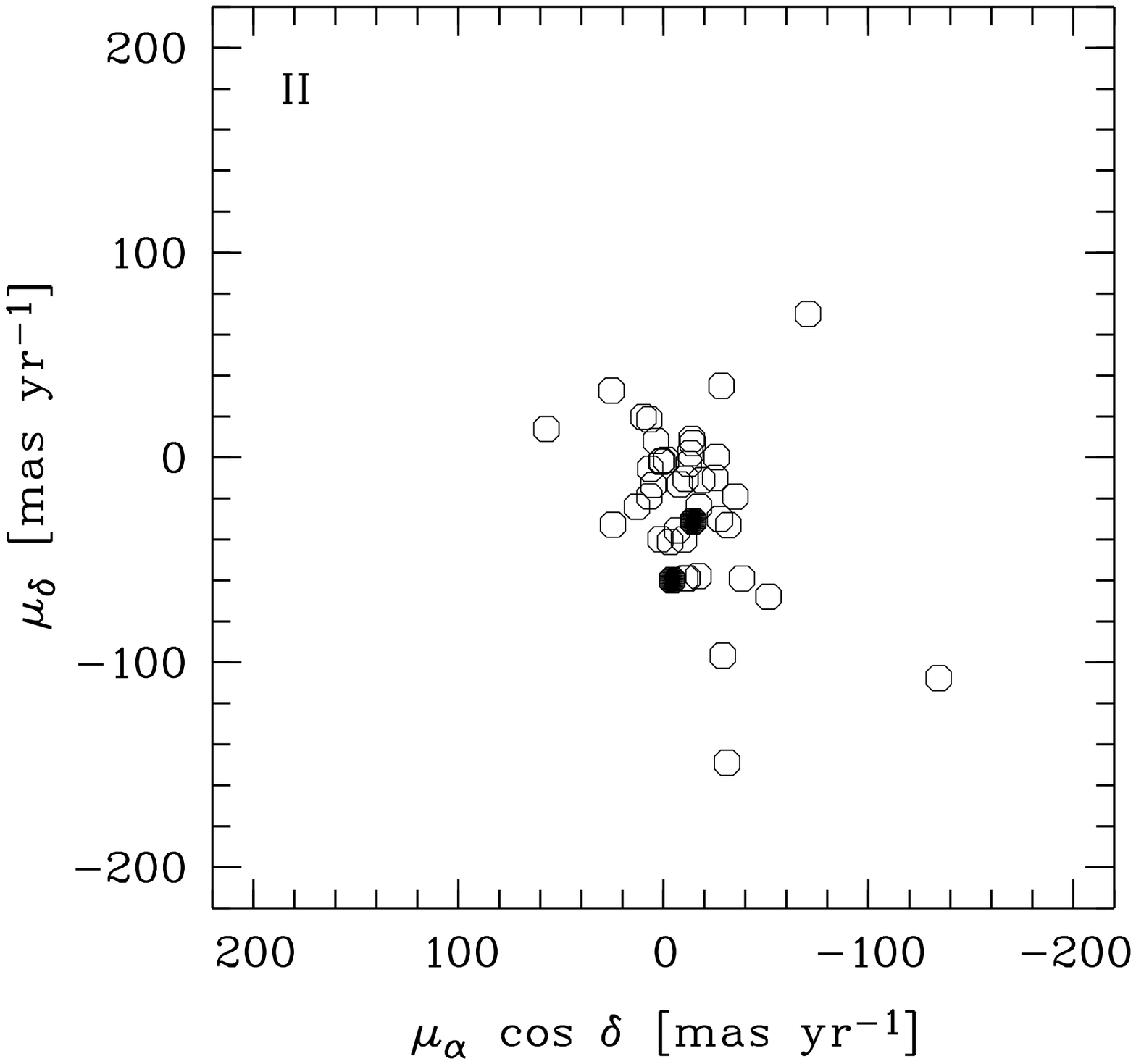}{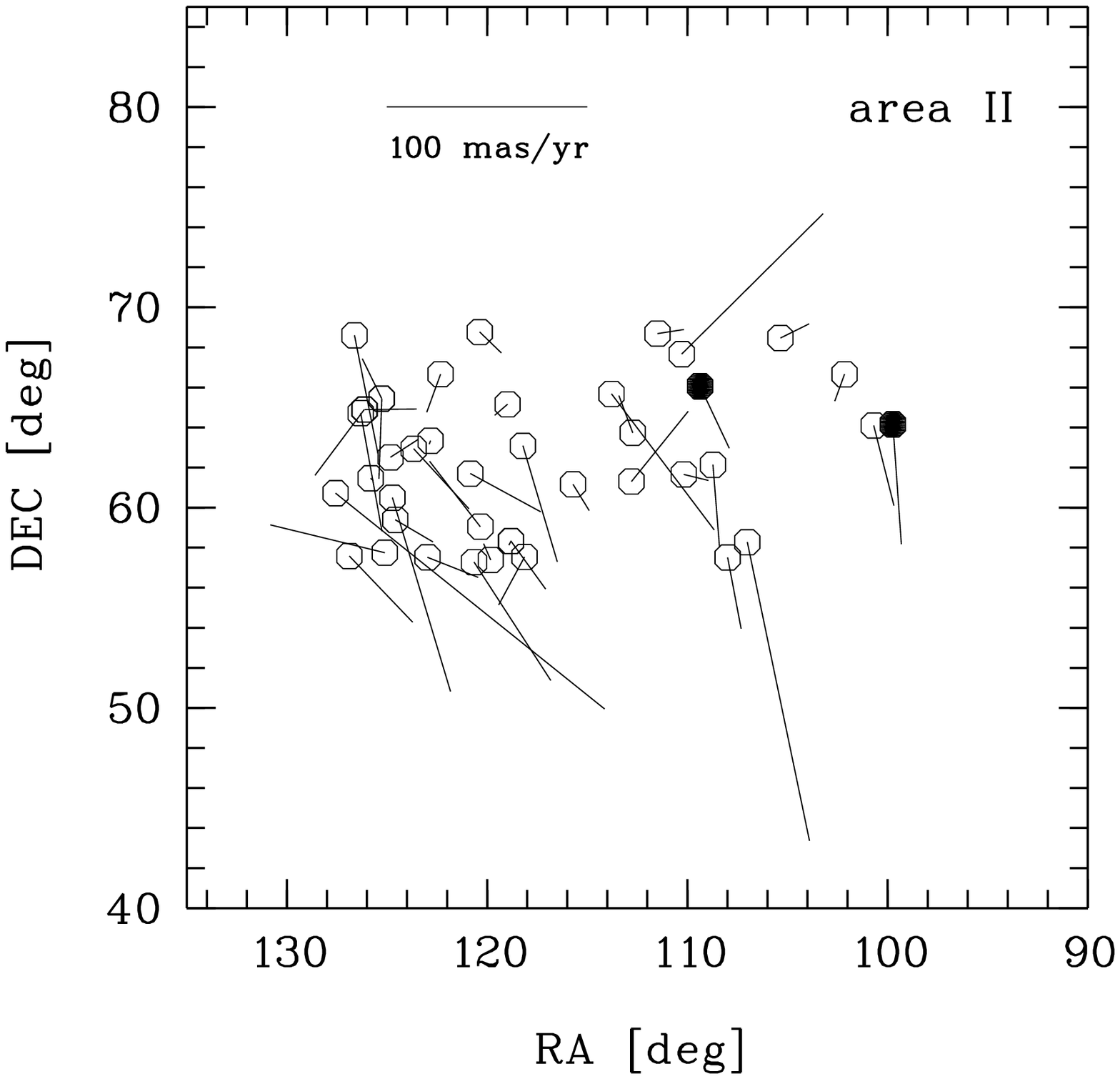}}\\
{\plottwo{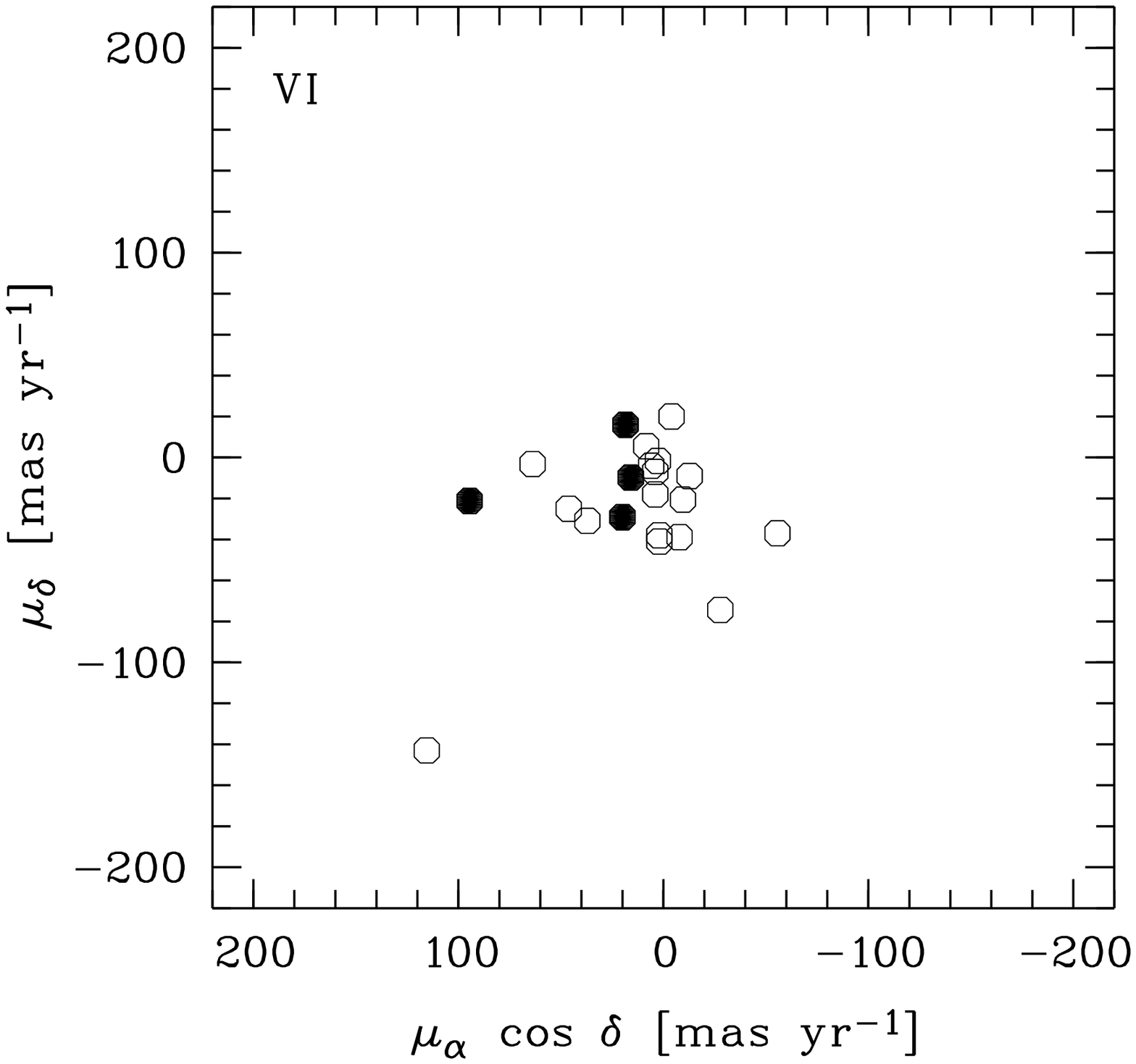}{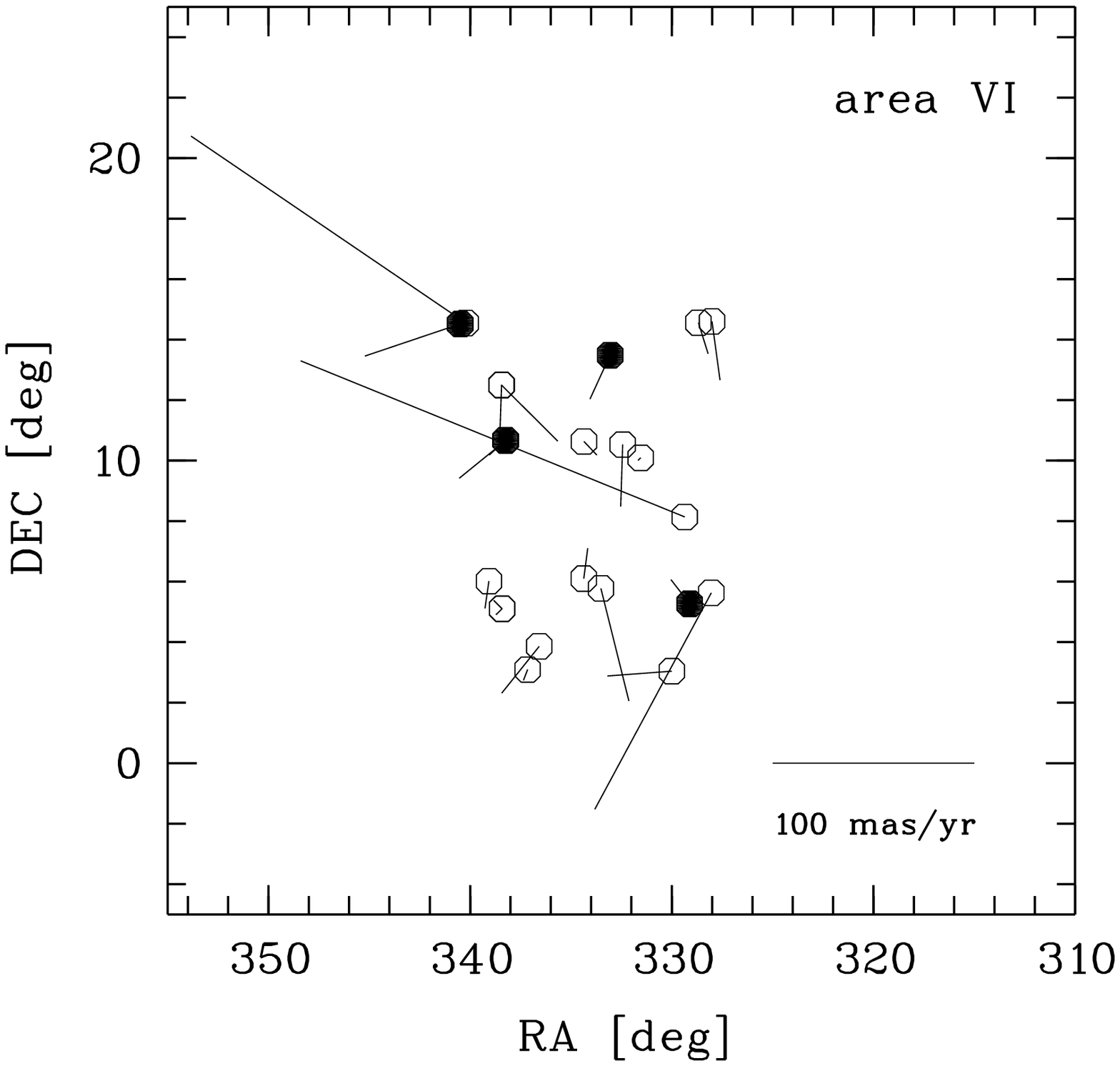}}
\caption[]{Proper motions and projected directions of motion for area II
and VI. Filled symbols: Li-rich stars. 
}
\label{oprop}
\end{figure}

The heliocentric radial velocities measured for stars in area I are shown in 
Fig. \ref{radvel}. For most Li-rich stars the radial velocity (RV) is close to 
the mean RV for the Taurus T\,Tauri stars of 16-17\,km\,s$^{-1}$. The mean and 
range of the RV measured by Neuh\"auser et al. (1997) 
for PMS stars south of Taur-Aur 
are displayed as long and short dashed lines, respectively. These stars are
probably located at a mean distance of $\sim 140$\,pc. The two stars with the
lowest $v_{\rm rad}$ are the same as mentioned above. 

\index{star|age}
\index{star|pre-main sequence}
Zickgraf et al. (1998) estimated an age of $\sim30$\,Myr for 
the lithium-rich stars in area I. The  
proper motions and radial velocities suggest that they
form a kinematically rather homogenous group consistent with their young age
(cf. also Frink et al. 1997).

Preliminary results of the investigation of proper motions in the other 
areas are shown in Fig.\ref{oprop}. In area II the Li-rich stars seem to 
follow the general trend in proper motion. In area VI three of the four 
Li-rich stars move in the same direction whereas for the other stars no 
clear pattern is discernible. 
The scatter of the proper motions is different for each study area. 
Areas II and V have the smallest scatter ($30-50$\,mas\,yr$^{-1}$), 
area III with $\sim150$\,mas\,yr$^{-1}$ the largest.

\begin{table}[tbh]
\caption[]{Lithium equivalent widths and abundances for G, K and M stars 
(on the usual scale of $\log N$(H) = 12) with W(Li)$>$50\,m\AA . 
Estimated uncertainties of the abundances are about 0.3-0.4\,dex.
Sequence number refer to Appenzeller et al. (1998).
F015 is a binary system. Both components show \ion{Li}{1} absorption. 
}
\large
\begin{tabular}{lllll}
\noalign{\smallskip} \hline \noalign{\smallskip} 
Sequ. $\#$ & ROSAT name & Sp. type &    $W$(Li)[m\AA ] & $\log N$(Li)\\
\noalign{\smallskip} \hline \noalign{\smallskip} 
A001  & RX\,J$0328.2+0409$ & K0   & 180  & +2.6 \\
A010  & RX\,J$0331.1+0713$ & K4e  & 410  & +2.9 \\
A042  & RX\,J$0339.9+0314$ & K2   & 140  & +2.0 \\
A057  & RX\,J$0344.4-0123$ & G9   & 290  & +3.1  \\
A058  & RX\,J$0344.8+0359$ & K1e  & 250  & +2.7 \\
A071  & RX\,J$0348.9+0110$ & K3(e)& 250  & +2.3 \\
A094  & RX\,J$0355.2+0329$ & K3e  & 350  & +2.7 \\
A095  & RX\,J$0355.3-0143$ & G5   & 230  & +3.1  \\
A100  & RX\,J$0358.1-0121$ & K4e  & 310  & +2.4 \\
A101  & RX\,J$0358.9-0017$ & K3e  & 265  & +2.4 \\
A104NE& RX\,J$0400.1+0818$ & G5   & 315  & +3.5  \\
A126  & RX\,J$0405.6+0341$ & G0   & 70  &  +2.7 \\
A144  & RX\,J$0412.1+0044$ & G0   & 60  & +2.6 \\
A151NW& RX\,J$0415.4+0611$ & G0   & 60  &  +2.6\\
A151SE& RX\,J$0415.4+0611$ & G0   & 80  & +2.7 \\
A161  & RX\,J$0417.8+0011$ & M0e  & 380  & +1.7 \\
B002  & RX\,J$0638.9+6409$ & K2   & 255  & +2.5 \\
B039  & RX\,J$0717.4+6603$ & K3e  & 200  & +2.2  \\
B206  & RX\,J$0828.1+6432$ & M1e  & 620  & +2.3  \\
C176  & RX\,J$1059.7-0522$ & K1   & 170  & +1.8 \\
C200  & RX\,J$1105.3-0735$ & K5e  & 100  & +1.3 \\
F015  & RX\,J$2156.4+0516$ & K2   & 140/190  & +2.0/+2.3  \\
F046  & RX\,J$2212.2+1329$ & G5   & 110  & +2.6  \\
F101  & RX\,J$2232.9+1040$ & K2   & 190  & +2.3 \\
F140  & RX\,J$2241.9+1431$ & K0   & 300  & +3.0 \\
\noalign{\smallskip} \hline \noalign{\smallskip} 
\end{tabular}
\label{litab}
\end{table}

%
%

%

\index{*V774 Tau|see HR\,1322}
\index{*V891 Tau|see HR\,1322}

\end{document}